\documentclass[twocolumn,showpacs,preprintnumbers,amsmath,amssymb,aps,prl]{revtex4}
\def\be{\begin{equation}}
\def\ee{\end{equation}}
\def\bea{\begin{eqnarray}}
\def\eea{\end{eqnarray}}
\def\nn{\nonumber}

\def\p{\partial}
\begin{document}

\preprint{hep-th/0601147~v3}
\title{\bf The first law of thermodynamics of the ($2+1$)-dimensional BTZ black holes and Kerr-de Sitter
spacetimes \footnote{This program was supported by a fund from Central China Normal University.}}
\author{WANG Shuang, WU Shuang-Qing \footnote{E-mail address: sqwu@phy.ccnu.edu.cn (Correspondent)},
XIE Fei, and DAN Lin}
\affiliation{\it\small College of Physical Science and Technology, Central China Normal University,
Wuhan, 430079}
\date{January 20, 2006}\revised{February 21, 2006}

\begin{abstract}
{\em\noindent
We investigate the first law of thermodynamics in the case of the ($2+1$)-dimensional BTZ black holes
and Kerr-de Sitter spacetimes, in particular, we focus on the integral mass formulas. It is found that
by assuming the cosmological constant as a variable state parameter, both the differential and integral
mass formulas of the first law of black hole thermodynamics in the asymptotic flat spacetimes can be
directly extended to those of rotating black holes in anti-de Sitter and de Sitter backgrounds. It
should be pointed that these formulas come into existence in any dimensions also.}
\end{abstract}

\maketitle

Since the seminal work of Bekenstein and Hawking, black hole thermodynamics$^{[1]}$ had been well
established over the past forty years via a close analogy to the four laws of the usual thermodynamical
system.$^{[2]}$ In particular, it is generally proved$^{[3]}$ that for asymptotic flat black holes in any
dimensions, the differential expression of the first law takes the form
\be
dM = TdS +\Omega dJ \, ,
\label{DMF1}
\ee
and the corresponding integral Bekenstein-Smarr mass formula is given by
\be
\frac{D-3}{D-2}M = TS +\Omega J \, ,
\label{IMF1}
\ee
where $D$ is the dimension of spacetime, $M$, $J$, $\Omega$, $T$, and $S$ are the ADM mass, angular
momentum, angular velocity, temperature, and Bekenstein-Hawking entropy of the black hole, respectively.

With the advent of AdS/CFT correspondence,$^{[4]}$ study of thermodynamical properties$^{[5]}$ of black
holes, especially in the background geometry with a nonvanishing cosmological constant, attracted considerable
attention in recent years.$^{[6-8]}$ For instance, it has been recently shown by Gibbons {\it et al}.$^{[6]}$
that with a suitable definition of the conserved charges such as the mass and angular momentum, the differential
mass formula (\ref{DMF1}) holds true for rotating black holes in anti-de Sitter backgrounds in four, five and
higher dimensions. However, under the assumption of an invariable cosmological constant, it is clear that the
integral expression (\ref{IMF1}) can never be satisfied any more. In order to rectify this situation, it is
possible for us to consider the cosmological constant $\Lambda =\pm (D-1)(D-2)/(2l^2)$ as a variable parameter
and promote it to a thermodynamic state variable,$^{[9]}$ so that the differential and integral mass formulas
can be modified to$^{[10]}$
\bea
&& dM = TdS +\Omega dJ +\Theta dl \, , \label{DMF2} \\
&& \frac{D-3}{D-2}M = TS +\Omega J +\frac{1}{D-2}\Theta l \, ,
\label{IMF2}
\eea
where $\Theta$ is the generalized force conjugate to the state parameter $l$.

In this Letter, we simply take the ($2+1$)-dimensional Banados-Teitelboim-Zanelli (BTZ) black hole$^{[11]}$
and Kerr-de Sitter spacetime$^{[12]}$ as explicit examples to illustrate the strategy that leads to the above
observation. A detailed study of this problem in arbitrary dimensions will be presented elsewhere.$^{[10]}$

First of all, let us discuss the first law of thermodynamics of the ($2+1$)-dimensional BTZ black holes,$^{[11]}$
which is an exact solution of Einstein field equation with a negative cosmological constant $\Lambda = -1/l^2$.
[Quasilocal thermodynamics of ($2+1$)-dimensional BTZ black holes had been investigated in Ref. [13] within the
Brown-York formalism.] Although a ``toy'' model in some respects, the BTZ black hole has stirred significant
interest by virtue of its connections with certain string theories and its role in microscopic entropy calculations.
Furthermore, this model has been proven to be an especially useful ``laboratory'' for studying quantum-corrected
thermodynamics.

In ($2+1$)-dimensions, the line element of the BTZ black hole solution is$^{[11]}$
\be
ds^2 = -\Delta dt^2 +\frac{dr^2}{\Delta} +r^2\Big(d\phi -\frac{J}{2r^2}dt\Big)^2 \, ,
\label{BTZm}
\ee
where the lapse function
\be
\Delta = -M +\frac{r^2}{l^2} +\frac{J^2}{4r^2} \, ,
\ee
in which two integration constants $M$ and $J$ denote the mass and angular momentum of the BTZ black holes,
respectively. (The unit $G_3 = 1/8$ is adopted through out this paper.)

The outer and inner horizons are given by the condition $\Delta = 0$, and read
\be
r_{\pm} = l\sqrt{M \pm\sqrt{M^2 -J^2/l^2}}/\sqrt{2} \, .
\ee
It is easy to see that the mass and angular momentum can be expressed in terms of the outer and inner horizons as
\be
M = \frac{r_+^2 +r_-^2}{l^2} \, , \qquad J = \frac{2r_+r_-}{l} \, .
\label{MA1}
\ee
The Bekenstein-Hawking entropy of the BTZ black holes is twice the perimeter of the event horizon,$^{[11]}$
\be
S = 2L = 4\pi r_+ \, ,
\label{EBTZ}
\ee
the surface gravity and angular velocity can be computed
\bea
\kappa &=& \frac{1}{2}\frac{d\Delta}{dr}\Big|_{r = r_+} = \frac{r_+^2 -r_-^2}{r_+l^2} \, , \nn \\
\Omega_+ &=& -\frac{g_{t\phi}}{g_{\phi\phi}}\Big|_{r = r_+} = \frac{J}{2r_+^2} = \frac{r_-}{r_+l} \, .
\eea

Now we apply the method in Ref. [5] to investigate the thermodynamics for the BTZ black holes. Making use of Eqs.
(\ref{MA1}) and (\ref{EBTZ}), we can get the Christodoulou-type mass formula$^{[14]}$
\be
M = \frac{S^2}{16\pi^2l^2} +\frac{4\pi^2J^2}{S^2} \, .
\label{CMF1}
\ee
Treating $l$ as an invariable constant and varying the mass formula (\ref{CMF1}) yield the conventional differential
expression of the first law of black hole thermodynamics
\be
dM = TdS +\Omega dJ \, ,
\ee
where the Hawking temperature $T = \kappa/(2\pi)$ and angular velocity are given by
\bea
 T &=& \frac{\p M}{\p S}\Big|_{J, l} =  \frac{S}{8\pi^2l^2} -\frac{8\pi^2J^2}{S^3}
= \frac{r_+^2 -r_-^2}{2\pi r_+l^2} \, , \nn \\
 \Omega &=& \frac{\p M}{\p J}\Big|_{S, l} = \frac{8\pi^2J}{S^2} = \frac{J}{2r_+^2}
= \Omega_+ \, .
\eea
However, taking into account the integral mass formula, it is not difficult to discover that the previously
obtained expression$^{[8]}$
\be
M = (1/2)TS +\Omega J \, ,
\ee
cannot fit to the generic formula in the $D = 3$ case. On the other hand, we observe that the combination
\be
TS +\Omega J = 2r_+^2/l^2 \, ,
\ee
doesn't vanish in the 3-dimensional case. This hints that the effect of the cosmological constant must
be considered.

It should be stressed that in the above discussion we have treated the cosmological constant as a fixed parameter;
however, it is possible to promote the cosmological constant to a thermodynamic state variable.$^{[9]}$ It is
necessary to point out that the assumption of a variable cosmological constant is also evidently supported by
some recent work.$^{[15]}$ Here and from now on, we shall assume the cosmological constant is variable. Under
this assumption, the differential and integral mass formulas for the first law of the BTZ black hole thermodynamics
receive a correction item related to the cosmological constant $l$
\bea
dM &=& TdS +\Omega dJ +\Theta dl \, , \label{DMF3} \\
0 &=& TS +\Omega J +\Theta l \, ,
\label{IMF3}
\eea
where $\Theta$ can be regarded as a generalized force as $l$ has the dimension of length and can be computed as
follows
\be
 \Theta = \frac{\p M}{\p l}\Big|_{S, J} = -\frac{S^2}{8\pi^2l^3} = -\frac{2r_+^2}{l^3} \, .
\label{GF}
\ee
It should be noted that the above expressions for the surface gravity, angular velocity and generalized force
follow directly from their thermodynamic definitions and coincide with those obtained by other method.$^{[8, 11]}$

Combining with the previous results obtained in the 4-dimensional case,$^{[9]}$ we deduce that the general formulas
for the first law of thermodynamics should be recast into the forms of Eqs. (\ref{DMF2}) and (\ref{IMF2}) in the
case of rotating black holes in anti-de Sitter backgrounds in three, four, five and higher dimensions.

Next, we would like to extend the above discussion to the case of the ($2+1$)-dimensional Kerr-de Sitter
spacetimes$^{[12]}$ with a positive cosmological constant $\Lambda = 1/l^2$, where the metric is still
described by the line element (\ref{BTZm}) where the lapse function is now replaced by
\be
\Delta = M -\frac{r^2}{l^2} +\frac{J^2}{4r^2} \, ,
\ee
in which two parameters $M$ and $J$ represent the ADM mass and angular momentum of the Kerr-de Sitter spaces,
respectively.$^{[16]}$

Analogous to the BTZ black hole case, two roots of the vanishing lapse function $\Delta = 0$ are denoted as
\be
r_{\pm} = l\sqrt{M \pm\sqrt{M^2 +J^2/l^2}}/\sqrt{2} \, .
\ee
where the radius $r_+$ is the only cosmological horizon, instead of a black-hole horizon of the ($2+1$)-dimensional
Kerr-de Sitter space.$^{[12]}$ Formally, let
\be
r_- = ir_{(-)} = il\sqrt{\sqrt{M^2 +J^2/l^2} -M}/\sqrt{2} \, .
\ee
so that one can pledge $r_-$ is a pure-imaginary number. Obviously, it is easy to obtain
\be
M = \frac{r_+^2 -r_{(-)}^2}{l^2} \, , \qquad J = \frac{2r_+r_{(-)}}{l} \, .
\label{MA2}
\ee
and the Bekenstein-Hawking entropy
\be
S = 4\pi r_+ \, .
\label{EKdS}
\ee
which is twice the perimeter of the horizon too. The surface gravity and angular velocity can be evaluated as before
\bea
\kappa &=& \frac{1}{2}\frac{d\Delta}{dr}\Big|_{r = r_+} = -\frac{r_+^2 +r_{(-)}^2}{r_+l^2} \, , \nn \\
\Omega_+ &=& -\frac{g_{t\phi}}{g_{\phi\phi}}\Big|_{r = r_+} = \frac{J}{2r_+^2} = \frac{r_{(-)}}{r_+l} \, .
\label{KO}
\eea

With the aid of Eqs. (\ref{MA2}) and (\ref{EKdS}), it follows immediately that the mass formula of the Kerr-de
Sitter space reads
\be
M = \frac{S^2}{16\pi^2l^2} -\frac{4\pi^2J^2}{S^2} \, .
\ee
Similar to the case of a negative cosmological constant, the expressions (\ref{DMF3}) and (\ref{IMF3}) are applicable
for the first law of thermodynamics of the Kerr-de Sitter space, where the generalized force $\Theta$ retains its
preceding form (\ref{GF}), while the Hawking temperature and angular velocity can be evaluated according to their
thermodynamical relations
\bea
 T &=& \frac{S}{8\pi^2l^2} +\frac{8\pi^2J^2}{S^3} = \frac{r_+^2 +r_{(-)}^2}{2\pi r_+l^2}
= -\frac{\kappa}{2\pi} \, , \nn \\
 \Omega &=& -\frac{8\pi^2J}{S^2} = -\frac{J}{2r_+^2} = -\Omega_+ \, .
\eea
It should be pointed out that both of them receive a minus sign relative to their geometric definitions (\ref{KO}).
Incorporating the effect of spacetime dimensions, these mass formulas can be commendably extended to the cases of
higher dimensional de Sitter spacetimes also.

In summary, we have investigated the thermodynamics of the ($2+1$)-dimensional BTZ black holes and Kerr-de Sitter
spacetimes by considering the cosmological constant as a variable state parameter. What more important is, we
have obtained various mass formulas, many of which can be directly extended to those of rotating black hole on
the background of anti-de Sitter and de Sitter spacetimes in any dimensions. We mention that the main results
obtained in this article are applicable not only for the ($2+1$)-dimensional case, but also for higher
dimensional case.

It is interesting to further extend the present investigation to the charged BTZ black hole case.$^{[11]}$ This
issue will be investigated in a future work.


\end{document}